\documentclass[prl,superscriptaddress,showpacs,keywords,reprint]{revtex4-1}
\usepackage{graphicx,amsmath,amssymb,amstext,amsfonts,color,bbm,dsfont,epstopdf,mathtools}
\usepackage[colorlinks,linkcolor=red,citecolor=blue,urlcolor=blue]{hyperref}
\usepackage[latin9]{inputenc}
\usepackage[normalem]{ulem}

\usepackage{times}
\usepackage{multirow}
\usepackage{rotating}
\usepackage{float}

\newcommand{\be}{\begin{equation}}
\newcommand{\ee}{\end{equation}}
\newcommand{\ben}{\begin{eqnarray}}
\newcommand{\een}{\end{eqnarray}}
\newcommand{\bes}{\begin{subequations}}
\newcommand{\ees}{\end{subequations}}
\newcommand{\bF}{\begin{figure}}
\newcommand{\eF}{\end{figure}}

\newcommand\dx{\mathrm{d} x}

\newcommand\im{\mathrm{i}}

\DeclareMathOperator{\po}{\mathcal{P}}
\DeclareMathOperator{\id1}{\mathds{1}}
\newcommand{\pTr}[2]{\mathrm{Tr}_{\mathrm{#1}}\left[ {#2} \right]}
\newcommand{\one}{\mathbbm{1}}

\newcommand\cc{\mathbb{C}}

\definecolor{jens}{rgb}{0,.8,.5}

\definecolor{mathis}{rgb}{.9,.0,.9}
\definecolor{berlin}{rgb}{0,.8,.6}
\newcommand{\ber}[1]{{\color{black} #1}} 
\definecolor{adrian}{rgb}{0,.4,.7}

\newcommand{\ket}[1]{\left|#1\right\rangle}

\newcommand{\vienna}{Vienna Center for Quantum Science and Technology, Atominstitut, 
TU Wien, Stadionallee 2, 1020 Vienna, Austria}
\newcommand{\dahlem}{Dahlem Center for Complex Quantum Systems, 
Freie Universit\"{a}t Berlin, 14195 Berlin, Germany}

\begin{document}

\title{Towards experimental quantum field tomography with ultracold atoms}

\author{A. Steffens}
\affiliation{\dahlem}

\author{M.\ Friesdorf}
\affiliation{\dahlem}

\author{T.\ Langen}
\affiliation{\vienna}

\author{B. Rauer}
\affiliation{\vienna}

\author{T. Schweigler\,}
\affiliation{\vienna}

\author{R.\ H{\"u}bener}
\affiliation{\dahlem}

\author{J.\ Schmiedmayer}
\affiliation{\vienna}

\author{C. A. Riofr\'io}
\affiliation{\dahlem}

\author{J.\ Eisert}
\affiliation{\dahlem}

\begin{abstract}
The experimental realisation of large scale many-body systems has seen immense progress in recent years, rendering full tomography tools for state identification
inefficient, especially for continuous systems. In order to work with these emerging physical platforms, new technologies for state identification are required.
In this work, we present first steps towards efficient experimental quantum field tomography. 
We employ our procedure to capture ultracold atomic systems using atom chips, a setup that allows for the quantum simulation of static 
and dynamical properties of interacting quantum fields. 
Our procedure is based on cMPS, the continuous analogues of matrix product states (MPS), ubiquitous in condensed-matter theory.
These states naturally incorporate the locality present in realistic physical settings and are thus prime candidates for describing the physics 
of locally interacting quantum fields.
The reconstruction procedure is based on two- and four-point correlation functions, from which we predict higher-order correlation functions,
thus validating our reconstruction for the experimental situation at hand.
We apply our procedure to quenched prethermalisation experiments for quasi-condensates.
In this setting, we can use the quality of our tomographic reconstruction as a probe for the non-equilibrium
nature of the involved physical processes.
We discuss the potential of such methods in the context of partial verification of analogue quantum simulators.
\end{abstract}
\date{\today}

\maketitle

Recent years have seen a rapid development in the field of quantum technologies: Complex quantum systems with many degrees of freedom can be controlled
with unprecedented precision, giving rise to applications in quantum metrology 
\cite{Wineland}, quantum information \cite{BlattErrorCorrection,Wineland}, and quantum simulation
\cite{BlochSimulation,Mimicking,CiracZollerSimulation}. 
This holds true specifically for architectures
based on trapped ions \cite{IonSimulation} and ultracold atoms \cite{BlochSimulation,Mimicking,LangenNature,Trotzky_etal12}, 
where large system sizes can now routinely be reached, while still maintaining control at the level of single constituents.
In the light of this development,  the mindset has been shifted when it comes to the assessment and verification of preparations of quantum states.
Traditionally, experiments are being used as a vessel to test theoretical models and descriptions by comparing their predictions to specific experimental output.
If approximate agreement is reached, this is then taken as evidence for the validity of the model at hand.

With quantum experiments of many degrees of freedom becoming significantly more accurate 
and with an attitude of ``quantum engineering'' and quantum simulation  taking over in this context, 
the standards in {\it quantum system identification} 
have similarly risen. 
Quantum state tomography \cite{QuantumTomography,Compressed,Compressed2,MPSTomo} fulfils this need for precise quantum state identification. 
It asks the question: Given data, what is the unknown quantum state compatible with those data? 
Compared to the traditional mindset, it is rather the converse approach, not predicting data from theoretical models, 
but rather retro-dicting the unknown physical situation at hand given data. 
This possibly more careful---and in a healthy way pedantic---mindset has done a good service in the past, 
and seems appropriate, given the developments in the precise control of quantum systems with many degrees of freedom.
Maybe unsurprisingly, the interest in the field of quantum system identification and specifically of quantum state tomography has exploded in recent years.

Yet, rather obviously, for quantum systems with many degrees of freedom, a new obstacle has to be overcome. 
Unqualified quantum state tomography must be inefficient in the system size, as exponentially many numbers need to be specified. 
For continuous systems of quantum fields, a priori it is not even clear what quantum field tomography should precisely mean. 
This observation has given way to the insight that in practical contexts, only those states need to be reconstructed that are encountered in natural contexts, 
as one only needs to be able to capture those states well. 
This insight has given rise to more economical and efficient notions of tomography, 
ranging from {\it quantum compressed sensing} \cite{QuantumTomography,Compressed,Compressed2} (to capture states of approximately low rank well), 
over {\it permutation-invariant tomography}, to {\it matrix-product state tomography} \cite{MPSTomo,MPOTomo,Efficient,Wick_MPS}. 
The latter approach is truly efficient for large systems and captures natural states well that exhibit 
low entanglement and a suitable decay of correlations. 
This is a perfectly meaningful approach, as long as one can give evidence that the state 
encountered is well-approximated by a representative in this class of states. 
In this sense, matrix-product states have been identified to be the right ``data set'' 
having the appropriate ``sparsity structure'' to capture quantum many-body systems 
-- and a large body of literature in the condensed matter context backs 
up this intuition of the ``physical corner of Hilbert space'' 
\cite{VerstraeteBig,EisertAreaLaws,MPSSurvey,2011AnPhy}.  

\subsection*{Quantum field tomography}
In this work, we introduce the notion of and at the same time experimentally apply key steps towards fully fletched quantum field tomography. 
The data set at the basis of this is the natural continuous analogue of matrix-product states, the continuous matrix-product states (cMPS) \cite{cMPS1,cMPS2,cMPS_Calc}. 
We present the mathematically precise details of the formalism to reconstruct an unknown cMPS from correlation data 
in the accompanying manuscript for this approach \cite{LongPaper}.
These tools are applied to data from ultracold atoms using atom chips, 
one of the most important platforms for quantum simulation of static and dynamic properties.

For this setup, we perform a proof-of-principle instance of unbiased quantum field tomography, 
without making assumptions on the underlying physical model 
or employing any effective description in terms of a free system.
Our approach is thus, in principle, capable of capturing general states of locally interacting models.
What is more, additional measurement data can be used to build trust in the reconstructed state and verify that cMPS
 indeed constitute a reasonable class of states for the setup at hand.
For this, the reconstructed state can be used to predict measurement results, which in turn can be checked experimentally.
In our case, we will use higher-order correlation functions for that purpose and find that they can be predicted by the
reconstructed cMPS to excellent accuracy.

Having set the scene, we now turn to a more specific description of the problem at hand. 
At the basis of the approach are translation invariant cMPS of one species of bosonic particles, defined by state vectors
\begin{equation}
\ket{\psi_{Q,R}}=\pTr{aux}{\po\textrm{e}^{\int_{0}^{L}\dx \left(Q\otimes\hat{\one}+R\otimes\hat{\Psi}^{\dagger}(x)\right)}}\ket{\Omega},\label{eq:cMPS:def}
\end{equation}
where the collection of field operators $\hat{\Psi}(x)$, $x\in[0,L]$, respect the canonical commutation relations
\begin{equation}
	[\text{\ensuremath{\hat{\Psi}}}(x), \text{\ensuremath{\hat{\Psi}}}^{\dagger}(y)] =\delta(x-y), 
\end{equation}
and $\left|\Omega\right\rangle$ is the vacuum state vector, 
$Q,R\in\cc^{d\times d}$ are matrices acting on an {\it auxiliary $d$-dimensional space}, 
and $L$ is the length of the closed physical system. 
In this expression $\mathcal{P}$ denotes the path ordering operator and $\mathrm{Tr_{aux}}$ traces out the auxiliary space.
The bond dimension $d$ takes the same role as the bond dimension for matrix product states \cite{VerstraeteBig,EisertAreaLaws,MPSSurvey,2011AnPhy}. 
Low entanglement states are expected to be well-approximated by cMPS of low bond dimension; in turn, for suitably large $d$, 
every quantum field state can be approximated.

\subsection{Experimental application: 1D condensate}

\begin{figure*}[t]
\includegraphics[width=1.7\columnwidth]{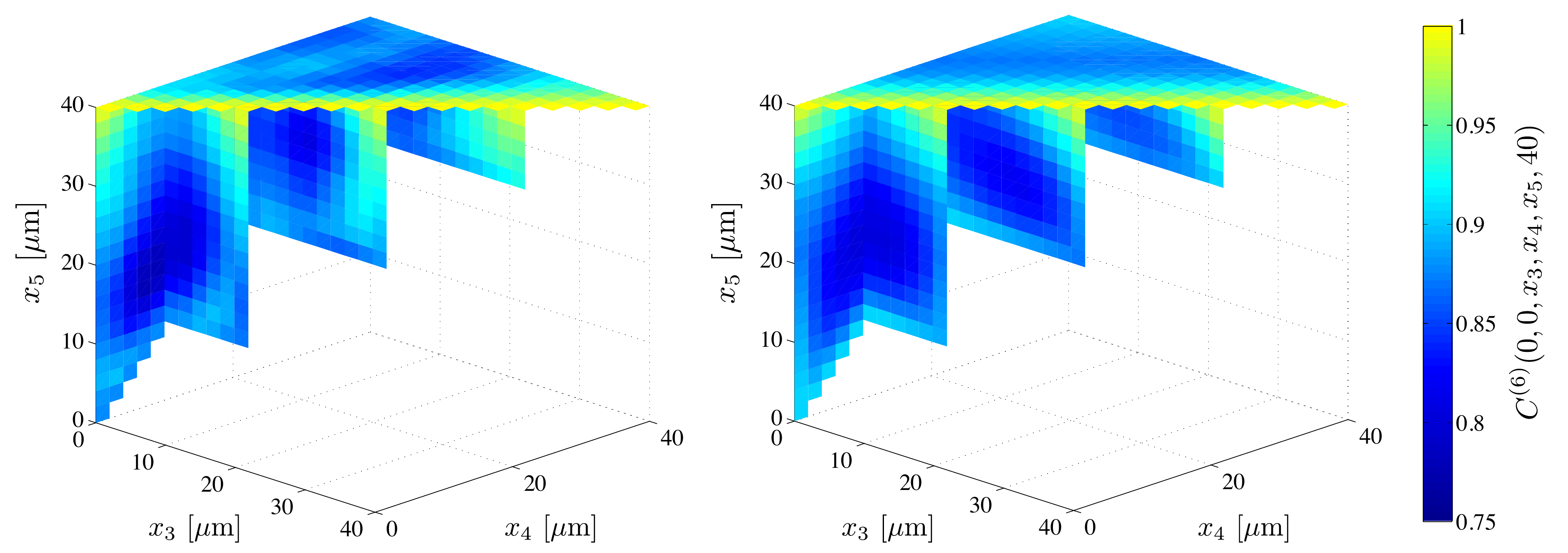}
\caption{Projections of the measured and predicted 6-point correlation function, measured (left) and reconstructed (right) 
 for a hold time after the quench of $t=3\text{ms}$. 
The great visual agreement between the experimental data and
the predicted correlation functions clearly demonstrates that the reconstruction of the full correlation behaviour of the state has been successful.}
\label{Experimental6pointFunction}
\end{figure*}

We employ our reconstruction procedure to perform quantum system tomography for a one-dimensional (1D) system of ultracold Bose gases, an architecture that in the
past has very successfully been used to test questions related to \emph{equilibration} and \emph{prethermalisation} \cite{Pretherm,LangenNature,SchmiedmayerScience}
and provides one of the prime setups for exploring the physics of interacting quantum fields.
The experiment consists of a large 1D quasi-condensate that is trapped using an atom chip.
To bring the system out of equilibrium, a split transversal to the condensate direction is performed, leading to two 1D condensates.
The setup in principle allows for different splitting procedures,
in particular an experimental scheme to test the Unruh effect with a 
specially modelled split has recently been proposed \cite{1402.6716}.
Once the split is complete, the condensates evolve independently and thus provide an ideal playground to understand equilibration and thermalisation 
for continuous quantum fields.
Finally, the phase correlations between the different condensates can be measured with a time-of-flight (ToF) expansion.
In all what follows, specifically the difference mode is being considered, which is 
\ber{nearly in the ground state before the quench and afterwards populated by a coherent superposition of phonon modes}
and thus \ber{still} approximately pure.
The subsequent out of equilibrium dynamics after the quench leads to apparent equilibration, 
prethermalisation and thermalisation. 
\ber{It} is important to stress that 
\ber{the evolution is unitary, leading to apparent equilibration}, 
perfectly compatible with the global state having little entropy \cite{CalabreseCardy06,CramerEisert08}, 
an insight that has been experimentally explored in the context of the systems 
considered here \cite{SchmiedmayerScience}.
Since the experimentally measured images are single shot measurements, 
repeating the experiment many times with identical initial conditions allows to measure 
not only the mean of the correlations, 
but also higher-order correlation functions are accessible \cite{full_distribution_function}.

The physics of this setup is well-approximated by the Lieb-Liniger model and its simplified low-energy approximation, 
the \emph{Tomonaga-Luttinger liquid} \cite{Takuya_NJP}.
The success in modelling the ground state of these systems with cMPS clearly suggests that 
they constitute a useful ansatz class to model the low-energy states of these locally interacting quantum fields
\cite{cMPS_Calc}.
While this is a clear indication that our method is ideally suited for the setup at hand, 
the employed reconstruction procedure is 
assumption-free and these theoretical models are not used in any way.
The systems trapped on the atom chip contain several thousand atoms and spread over sizes as large as $100 \, \mu \text{m}$.
In the middle of the trap, the system can be well-approximated by a homogeneous, translationally invariant quantum field,
allowing us to efficiently model the difference mode 
as a cMPS and to apply our reconstruction tools to partly recover 
the non-equilibrium state of the system \cite{LongPaper}.
This reconstruction is based on low-order correlation functions, 
which are sufficient to already obtain important structure of the cMPS.
In particular, the reconstructed information can be used to \emph{predict} higher-order correlation functions, 
which allows us to certify that the modelling of the state was successful 
and that the cMPS approach is indeed suited for the setup at hand.

In this proof-of-principle application, we specifically model novel data from prethermalisation experiments \cite{LangenNature,SchmiedmayerScience}, 
focussing on the state after an evolution time of $t = 3~\text{ms}$. 
While the tomography procedure for this short evolution time works extremely well, 
we find that the quality of a reconstruction with fixed bond dimension decreases substantially with increasing evolution time.
This drop in quality can be seen as a probe for the non-equilibrium nature of the quenched setting at
hand and provides a promising gateway towards testing the entanglement growth in these continuous systems.

Note that, even though the physical system at hand can be well captured with a free Tomonaga-Luttinger liquid model
\cite{Takuya_NJP}, the states of the system can still be strongly entangled, 
in the sense that entanglement entropies across any real space cut of the system are, in principle, arbitrarily large.
It is precisely this spatial entanglement that 
influences the quality of tensor network descriptions of the state and that is a key factor for the quality
of any cMPS reconstruction \cite{LongPaper}.

\subsection{Data analysis}

The phase correlation functions extracted from ToF images take the following form
\begin{equation*}
C^{\left(n\right)}\left(x_{1},\dots,x_{n}\right)
= \text{Re} \left\langle 
\textrm{e}^{\im\left(\hat{\theta}_{x_{1}}-\hat{\theta}_{x_{2}}+\hat{\theta}_{x_{3}}-\dots+\hat{\theta}_{x_{n-1}}-\hat{\theta}_{x_{n}} 
\right)}\right\rangle ,
\label{eq:applic:bec:phase expval}
\end{equation*}
where $\hat{\theta}_x$ is the phase difference operator of the two BECs and the angular brackets denote the ensemble average. While, in principle, the reconstruction of any correlation function is possible using the presented scheme, we restrict ourselves to even-order correlation functions, because only they can be measured in the experiment (see Appendix B).

To make the correlation function directly accessible to our reconstruction procedure, we must write it in terms of field operators $\hat{\psi}(x)$. 
For this purpose, we use the fact that $\hat{\theta}(x)$ commutes for different positions 
and employ the polar decomposition to construct an effective field operator
$\hat{\psi}(x)=\hat{n}(x)^{1/2} \textrm{e}^{-\im\hat{\theta}(x)}$,
where $\hat{n}(x) = {\hat{\psi}^{\dagger}(x)\hat{\psi}(x)}$ is taken to be the density of one of
the two condensates. The construction ensures that these effective field operators indeed fulfil the correct commutation
relations.
We can write this $n$-point function in terms of field operators
\begin{equation}\label{eq:ExperimentCorrelation}
  C^{\left(n\right)}\left(x_{1},\dots,x_{n}\right)
  =\left\langle \hat{n}(x_{1})^{-\frac{1}{2}}\hat{\psi}^{\dagger}(x_{1})\hat{\psi}(x_{2}) \hat{n}(x_{2})^{-\frac{1}{2}}\dots\right\rangle.
\end{equation}
Note that this corresponds to employing an effective modelling of the two condensates as one continuous quantum field, which captures their phase 
difference \cite{Takuya_NJP}.
Since it is sufficient for performing the tomography procedure, 
we will restrict the correlation information to the normal ordered set with 
$x_1 \leq x_2 \leq \cdots \leq x_n$.
The density operator involved can be taken to be the density of one of the two condensates. 
In the cMPS language, assuming translation invariance and the thermodynamic limit, this can be reformulated as 
\begin{equation}\label{eq:cor_fun}
C^{(n)}\left(\tau_{1},\dots,\tau_{n-1}\right)= 
   \sum_{\{k_j\}}^{d^{2}}  \rho_{k_1,\dots,k_{n-1}}\textrm{e}^{\lambda_{k_{1}}\tau_{1}}\dots\textrm{e}^{\lambda_{k_{n-1}}\tau_{n-1}}
\end{equation}
where $\tau_{k}=x_{k+1}-x_{k}$, 
\begin{equation}\label{eq:residues}\rho_{k_1,\dots,k_{n-1}}=M_{1,k_{n-1}}^{-1}M_{k_{n-1},k_{n-2}}\dots M_{k_{2},k_{1}}^{-1}M_{k_{1},1},
\end{equation}
the $\lambda_k$ are the eigenvalues of the transfer matrix $T$, 
and $M$ is $\overline{R}{}^{\frac{1}{2}}\otimes R^{-\frac{1}{2}}$ in the diagonal basis of $T$~\cite{cMPS1} (see Appendix C). 
Having established the proper model for the experimental data, we proceed to applying our reconstruction theory. 
Here, we will follow the more mathematically minded outline of Ref.\ \cite{LongPaper}, where only a few adjustments to the general treatment are needed.

\begin{figure*}[t]
\includegraphics[width=1.7\columnwidth]{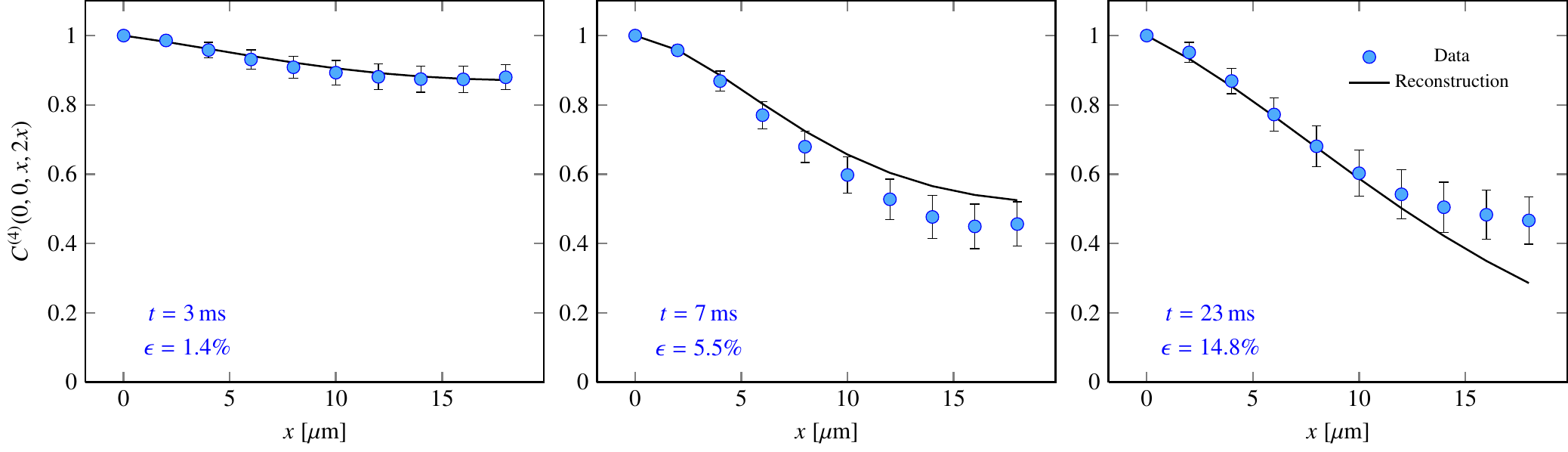}
\caption{Projection of the four-point correlation function 
  for different hold times after the quench in the prethermalisation experiment. 
  The quality of the cMPS ansatz decreases substantially with the hold time (see main text).
  The error measure $\epsilon$ is the mean relative deviation of the full four-point correlator
  (see Appendix D). 
}
\label{2d_Cut}
\end{figure*}

The goal of our tomography procedure is to partially reconstruct the experimental state of the system.
The great advantage of cMPS as an ansatz class is that the knowledge of two- and four-point correlation functions is already sufficient for this task,
rendering it feasible even for the continuous large scale many-body system at hand. 
The reconstruction proceeds by first extracting the eigenvalues $\lambda_k$ from the two-point correlation function, 
which was performed by a least-squares fit and under the assumption of translational invariance for the modelled system in the middle of the trap. 

In a second step, a compatible $M$ matrix is determined. 
For this, we rely on the explicit calculation of the correlation function from the $M$ matrix and the eigenvalues
of $T$ according to Eqs. \eqref{eq:cor_fun} and \eqref{eq:residues}.
Using the fitted two-point correlation function together with a normalisation condition, some entries of $M$
can already be fixed \cite{LongPaper,Wick_MPS}. The remaining entries are obtained by a Nelder-Mead simplex algorithm 
that varies them and compares the predicted four-point correlators. 
The optimisation was fully implemented in python relying on numerical tools from the open source Scipy library \cite{scipy}.
Working with real $M$ matrices of bond dimension $d=2$ and relying on a set of 100 random initial positions 
proved to be sufficient for approximating the measurement data well.
To estimate how well the reconstruction of the four-point correlation function worked, we use the mean relative deviation
(see Appendix D),
and find a small error of 1.4\%, which is of the same magnitude as the experimental errors.

Naturally, approximating a correlation function can be done in many ways and it is, a priori, not clear that one has truly
gained knowledge about the state. The advantage of the cMPS ansatz class is that the approximation performed above is sufficient to fully reconstruct the correlation behaviour of the cMPS.
This can be used to build trust in the reconstructed state by using it to predict higher-order correlation functions,
  which in turn can be experimentally checked.
This provides an excellent benchmark for our procedure and allows us to estimate the quality of our guess for the unknown experimental state.

In Fig.~\ref{Experimental6pointFunction}, we show projections of the relevant sections of the experimental and predicted
six-point function.
This image shows some of the \emph{volumetric} elements of the projections 
of the high-dimensional six-point correlation function array and
demonstrates a great overall agreement between experimental data and the \emph{predicted} correlation data.
More quantitatively, as a figure of merit for measuring the performance of the reconstruction, 
we use the mean relative deviation 
over all indices belonging to the relevant simplex of the data with $x_1\leq x_2\leq \dots \leq x_n$
(see Appendix D). 
We find that for the six-point function the mean relative deviation is 2.1\% and the maximum relative deviation is 10.3\%. 
Both variations are in the same order of magnitude as the expected experimental measurement error.
This benchmark clearly shows that the reconstruction of the full correlation behaviour of the state was successful,
providing a proof-of-principle application for efficient state tomography of interacting many-body quantum fields.

The method presented here provides a possibility to efficiently reconstruct low-energy states 
of a locally interacting quantum fields, a setting where no other reconstruction procedure is applicable.
Moreover, steps can be taken to verify the quality of the reconstruction, 
thus allowing for an a posteriori verification that
the employed method is suitable to describe the physical setting at hand.

Naturally, while quantum field tomography necessarily has to rely on a finite-dimensional
``data set'', it is clear that not all situations can be equally well captured by the approach proposed here. 
This method applies to states of low entanglement, a situation expected to be present for
ground states or states in non-equilibrium following quenches for short times.
It will surely be difficult to capture highly entangled or thermal states, 
which are expected to have a high description complexity, with these tools \cite{LongPaper}: 
This leads to the intriguing situation that those states that cannot be reconstructed are also those states that 
can not be classically simulated with variants of the density matrix renormalisation group approach, relying on MPS or cMPS. 

Interestingly, we observe that the quality of our reconstruction decreases substantially with increased hold
time after the quench (see Fig.~\ref{2d_Cut} and Appendix D).
This can be seen as an indicator for the non-equilibrium phenomena associated with this quenched system.
There are several possible explanations for the decrease in quality that we observe.

The physics of sudden quenches in discrete settings is usually connected to a linear entanglement growth
with time \cite{SchuchQuench,CramerEisert08,EisertAreaLaws}, 
while for each time satisfying an area law in space \cite{EisertAreaLaws}.
A similar behaviour in the continuous setting at hand would be a possible explanation 
for the drop in  quality that we observe. 
Since our reconstruction with a bond dimension $d=2$ cMPS is only well-suited for states
with low entanglement, one naturally expects it to fail for long times when entanglement has been 
build up.
Indeed, such light cone dynamics, connected to the growth of spatial entanglement in real space 
in continuous systems has recently been made explicit experimentally in Refs.~\cite{LightConeSchmied,LangenNature}.

There are also other factors that may influence the quality of the reconstruction we perform. 
Experimental imperfections or the remaining actual temperature 
in the system \ber{can lead} to a true mixedness of the state,
and it is not quite clear to what extent the pure-state prescription is still fully appropriate.
Previous studies based on pure-state Luttinger liquids provide strong evidence that the system is still close to being pure even for
long evolution times \cite{LightConeSchmied}.
Another potential factor is that the experiment as it is done with present technology 
necessarily takes place in a trap.
The experimental data was taken in the middle of the trap, where, initially, the assumption of translational invariance
holds up to excellent accuracy. 
For long hold times after the quench, regions outside of the center of the trap will, however,
have an influence on the behaviour of the system in the middle. 
It is, for example, known that the characteristic speed of
sound has to be altered in the presence of a trap in these continuous systems \cite{LangenNature}.

We have confirmed that the qualitative behaviour of a decreasing approximation quality for longer hold times
also occurs for other data sets corresponding to different initial states.
It seems an exciting further perspective to quantitatively explore the entanglement behaviour in time 
and the accompanied description complexity, 
and to conclusively discriminate it from undesired experimental influences.

\subsection*{Further perspectives and partial verification tools\\ for quantum simulators}
In this work, we have demonstrated a first proof-of-principle application for continuous interacting quantum fields, a setup that has already provided
important insights into the physics of equilibration and thermalisation in the past.
The experimental data was sufficient for reconstructing the full correlation information of the state, which allowed us to benchmark our results.
This was done by predicting higher-order correlation functions from the reconstructed state, which showed excellent agreement with the corresponding
measured quantities.
What is more, we observed that the fits become increasingly difficult for larger evolution times, which is reminiscent 
of numerical problems connected to the linear growth of entanglement after sudden quenches in discrete systems. This surely is merely a first step in the direction of a
larger programme. Still, in this work, we advocate a mild {\it paradigm change} in the reconstruction of quantum fields: Instead of trying to set up a 
model and to compare predictions of the model with data, one puts the data into the focus of attention and attempts a reconstruction in the mindset of quantum tomography.

Such a partial verification scheme also provides an interesting perspective for the field of quantum simulations in general:
In many important applications, it seems likely to find the final result of a simulation in some ``physical corner'' \cite{VerstraeteBig,EisertAreaLaws}
of the full Hilbert space, for example
parametrised by cMPS (or MPS). This holds true even if the evolution is highly non-trivial and explores a large state space.
A particular setting that comes to mind is given by thermalisation or open-system dynamics, where the final states are at least 
conjectured to be known, yet the time scales to achieve them are inaccessible with current theoretical methods.
In these settings, tomography tools, such as the one presented in this work, would not only allow to read-out the result of the quantum simulation,
but would further provide meaningful ways to cross-check the simulation data and perform a partial validation, even if 
numerical simulations of the full evolution are impossible to perform.
Such a partial validation scheme seems to be a prime candidate to build trust in otherwise completely inaccessible results of quantum simulations
and is therefore a highly intriguing prospect.

\appendix
\subsection{Appendix A: Experimental considerations}
In the experiment, a single specimen of an ultracold gas of ${}^{87}$Rb atoms is prepared using evaporative cooling on an atom chip. The final temperature and the chemical potential of the gas are both well below the first radially excited state of the trapping potential, implementing 
a one-dimensional bosonic system that is well-approximated by the Lieb-Liniger model. A 
sudden global quench is realised by transversally splitting the gas into two mutually coherent halves, 
leading to an out of equilibrium, approximately pure state.
Subsequently, this non-equilibrium system is let to evolve in the trap for a variable hold-time. Its dynamical states are probed using matter wave interferometry in time-of-flight, which enables the direct measurement of the local relative phase $\theta(x)$. The corresponding correlation functions are constructed by averaging over approximately $150$ experimental realisations.

\subsection{Appendix B: Accessible correlation functions}
We are restricted to even-order correlation functions in the experiment. The reason for this is the fact that many experimental realisations are needed to construct the correlation functions. Each of these experimental realisations provides us with a measurement of the relative phase 
\begin{equation}
	\theta(x) = \phi(x) + \varphi. 
\end{equation}
Here, $\phi(x) $ is the actual fluctuating phase that contains the interesting many-body physics 
and $\varphi$ is a small global phase diffusion that is random in every experimental realisation \cite{Schumm05}.  
This global phase diffusion results from small shot-to-shot fluctuations in the electrical currents that create the trapping potential. 
These cause small random imbalances of the double well, leading to random and unknown values for $\varphi$. 
For the even-order correlation functions only differences between the $\theta$ at different positions need to be evaluated. 
Consequently, the global shifts $\varphi$ cancel automatically. 
However, for odd-order correlation functions contributions $\sim e^{i\varphi}$ remain. 
Hence, the measured result does not only contain the pure dynamics, 
but is significantly perturbed by the unknown fluctuations of $\varphi$.

\subsection{Appendix C: Correlation functions and cMPS}
As discussed 
in the main text, our reconstruction methods are based on the cMPS formalism, building upon state vectors that can be written in the form
\begin{equation}
\ket{\psi_{Q,R}}=\pTr{aux}{\po\textrm{e}^{\int_{0}^{L}\dx \left(Q\otimes\hat{\one}+R\otimes\hat{\Psi}^{\dagger}(x)\right)}}\ket{\Omega}.
\end{equation}
This formalism provides us with an efficient way to compute correlation functions for quantum fields. 
The correlation functions in Eq.\ \eqref{eq:ExperimentCorrelation} can be directly calculated in terms of the cMPS variational parameter matrices $R$ and $Q$ in the thermodynamic limit as
\begin{align}\label{eq:corrfunc_exp}
  \nonumber
  C^{\left(n\right)}\left(x_{1},\dots,x_{n}\right)
  =\mathrm{Tr} \Big ( \overline{R}^{\frac{1}{2}}&\otimes R^{-\frac{1}{2}}
  \textrm{e}^{T\tau_{1}}
  \overline{R}^{-\frac{1}{2}}\otimes R^{\frac{1}{2}}
  \textrm{e}^{T\tau_{2}} \dots\\
  \dots\overline{R}^{\frac{1}{2}}\otimes R^{-\frac{1}{2}}
  \textrm{e}^{T\tau_{n-1}}
  \overline{R}^{-\frac{1}{2}}&\otimes R^{\frac{1}{2}}
  \lim_{L\rightarrow\infty}\textrm{e}^{T(L-x_{n})} \Big )
\end{align}
with the transfer matrix 
\begin{equation}	
	T:=\overline{Q}\otimes\id1_d+\id1_d\otimes Q+\overline{R}\otimes R,
\end{equation}
and positive distances $\tau_{j}=x_{j+1}-x_{j}$ for $j= 1,\dots,n-1$ and $\tau_{n}=L-x_{n}$. 
The overline denotes the complex conjugation. We arrive at Eq.~\eqref{eq:corrfunc_exp} by the 
correspondences between field operators and variational matrices as described in Ref.\ \cite{cMPS1}.
Eq.\ \eqref{eq:cor_fun} follows directly by writing all the matrices in the basis where the transfer matrix $T$ is diagonal. 
Specifically, the matrix $M\in\cc^{d^2\times d^2}$ is defined as 
\begin{equation}
	M=X^{-1}\left( \overline{R}\otimes R\right) X, 
\end{equation}
where $X$ is the matrix defined in the way that 
$X^{-1}TX$ is diagonal and compatible with the ordering of the eigenvalues $\{\lambda_k\}$.

\subsection{Appendix D: Error analysis}
To quantify the error of our tomography procedure, 
we use the relative mean deviation with respect to the fitted (reconstructed) data,
defined as 
\begin{align}
  \epsilon := \frac{1}{N} \sum_{\textbf{x} \in X} 
  \frac{|C(\textbf{x}) - C_\text{rec}(\textbf{x})|}{|C_\text{rec}(\textbf{x})|},
\end{align}
where $X$ is the set of all $\textbf{x} = (x_1, \dots, x_n)$, 
with $x_1 \leq x_2 \leq \cdots \leq x_n$ and $N$ denotes the number of discrete data points included.
The maximum relative deviation is defined as the largest summand.

We observe an increase of the relative mean deviation with the hold time after the quench (see Fig.\ \ref{different_times}).
As discussed in the main text, this can be seen as a clear signature of the non-equilibrium nature
of the physical processes involved.

\begin{figure}[H]
\centering
  \bgroup
  \def\arraystretch{1.4}
  \begin{tabular}{cr|rrr}
    \multicolumn{2}{c}{} & \multicolumn{3}{c}{\textit{Correlation error}} \\
    \multicolumn{2}{c}{} & $C_2$ & $C_4$ & $C_6$\\
    \cline{2-5}
    \multirow{3}{*}{\begin{sideways} \textit{Hold time}~ \end{sideways}} 
    &3 ms \hspace{.2cm} & \hspace{.2cm} 0.3\% & 1.4\% & 2.1\%\\
    &7 ms \hspace{.2cm} & 0.7\% & 5.5\% & 6.5\%\\
    &23 ms \hspace{.2cm} & 4.6\% & 14.8\% & 15.7\%\\
  \end{tabular}
  \egroup
  \caption[Error for different times]{Relative mean deviation  
    for the fitted two- and four-point correlators and the 
    reconstructed six-point correlator for different hold times after the quench.
    The tomography procedure works very well for short times, but is less accurate for longer quench times
    (see discussion in the main text). 
  }
  \label{different_times}
\end{figure}

\subsection*{Acknowledgements}
We thank the EU (SIQS, RAQUEL, COST), the ERC (TAQ and QuantumRelax), the Studienstiftung des Deutschen Volkes, the FQXi, and the BMBF (QuOReP) for support. 
B.R., T.S. and T.L. 
acknowledge support by the Austrian Science Fund (FWF) 
through the Doctoral Programme CoQuS (W1210).
We thank the KITP in Santa Barbara for hospitality and
acknowledge R.\ Geiger and A.\ Werner for useful discussions.

\bibliographystyle{utphys}

\providecommand{\href}[2]{#2}\begingroup\raggedright\endgroup

\end{document}